\documentclass[twocolumn]{aastex631}

\usepackage{graphicx}
\usepackage{amsmath}
\usepackage{soul}
\usepackage{xfrac}
\usepackage{cases}
\usepackage{changepage}

\def\equationautorefname~#1\null{equation~(#1)}

\DeclareMathAlphabet{\mathpzc}{OT1}{pzc}{m}{it}
\definecolor{purple}{RGB}{160,32,240}

\newcommand{\Msun}{M_{\odot}}

\newcommand{\Mj}{M_{\rm J}}

\newcommand{\Lsun}{L_{\odot}}
\newcommand{\Rj}{R_{\rm J}}

\newcommand{\Rp}{R_{\rm p}}

\newcommand{\Mdot}{\dot{M}_{\rm p}}
\newcommand{\Mp}{M_{\rm p}}

\shorttitle{Protoplanet atmospheres}
\shortauthors{Choksi et al.}

\begin{document}

\title{The accretionally heated atmospheres of protoplanets and brown dwarfs}

\correspondingauthor{Nick Choksi}
\email{nchoksi@caltech.edu}

\author[0000-0003-0690-1056]{Nick Choksi}
\affiliation{Division of Geological and Planetary Science, California Institute of Technology, Pasadena, CA 91125, USA}

\author[0000-0002-8518-9601]{Peter Gao}
\affiliation{Earth \& Planets Laboratory, Carnegie Institution for Science, Washington, DC, USA}

\author[0000-0002-6618-1137]{Jerry W. Xuan}
\affiliation{Department of Physics, University of California, Santa Barbara, Santa Barbara, CA, USA}

\author[0000-0001-6396-8439]{William Balmer}
\affiliation{Department of Physics \& Astronomy, Northwestern University, Evanston, IL, USA}

\author[0000-0001-5864-9599]{James Mang}
\affiliation{Department of Astronomy, University of Texas, Austin, TX, USA}


\begin{abstract}
The ranks of accreting planets and brown dwarfs are growing. To help interpret their spectra, we study how accretion affects a planet's atmosphere. The atmosphere is irradiated from above by whatever nonthermal radiation from the accretion shock does not directly escape into space. This accretional heating drives a temperature inversion. Even relatively slow accretion, corresponding to mass doubling times longer than system ages, can raise temperatures by many hundreds of Kelvin at ambient pressures of order millibars. The heated atmosphere radiates an optical/infrared excess. For environments that are not too dusty, the excess concentrates in molecular lines that trace the heated layer. Thus an accretionally heated atmosphere may be distinguished by its muted absorption bands. Irradiation from the accretion shock also changes the chemical composition of the atmosphere, further altering spectra. Atmospheric retrievals promise to map out the deposition of accretional energy, helping to decide between ``hot'' and ``cold'' start modes of planet formation. 
\end{abstract}


\section{Introduction}
Observations of 
accreting planets are providing a new and direct window into the planet formation process. The list of these confirmed ``protoplanets'' is short but growing. The first detections came in the PDS 70 system \citep{keppler_etal_2018, haffert_etal_2019, wang_etal_2020}, which hosts two super-Jupiters inside the cavity of a transitional disk. Adding to their ranks are the recently discovered  WISPIT 2b \& c (the former is the first planet detected inside of an annular disk gap of the kind classically predicted by theory; \citealt{vancap_etal_2025, close_etal_2025_wispit, lawlor_etal_2026}). All four of these protoplanets are a few million years old, weigh a few Jupiters apiece, and orbit at separations of tens of AU inside of larger, gaseous circumstellar disks. 
With a handful of tentative detections awaiting confirmation \citep[][]{currie_etal_2022, wagner_etal_2023, bernardi_etal_2026}, plus dozens of gapped circumstellar disks promising to harbor even more  \citep[e.g.][]{huang_etal_2018, oberg_etal_2021}, the ranks of protoplanets are poised to grow. 

To use these discoveries to learn about the planet formation process, we need to know how to interpret a protoplanet's spectrum. The total power released by accretion is $G\Mp\Mdot/\Rp$, where $G$ is the gravitational constant and $\Mp$ and $\Rp$ are the planet mass and radius. At what wavelengths does the accretion luminosity emerge? What physical processes and spatial scales are traced by different portions of the spectrum? 

Past studies have calculated the broadband spectral energy distributions of accreting circumplanetary disks or envelopes filling a planet's gravitational sphere of influence \citep{zhu_2015, szulagyi_etal_sed2, choksi_chiang_2022, choksi_chiang_2025, adams_batygin_2022, taylor_adams_2024, aster_etal_2026}. These extended, cooler structures produce excess emission redward of a few micron. Indeed, \textit{ALMA} imaging reveals mm-wave emission from PDS 70c \citep{isella_etal_2019, benisty_etal_2021} that can be fit by a circumplanetary disk \citep{shibaike_mordasini_2024, choksi_chiang_2025}. On smaller scales, the accreting gas shocks against the planet's hydrostatic atmosphere. The shocked gas reaches temperatures $T \sim 10^5$ K and radiates short-wavelength Balmer line and recombination continuum photons \citep{aoyama_etal_2020, aoyama_etal_2021}, as observed in both the PDS 70 and WISPIT systems \citep{haffert_etal_2019, close_etal_2025_wispit}.

Our focus in this paper lies in-between these short and long-wavelength regimes, in the optical/near-infrared, where the bulk of the luminosity of known protoplanets is emitted  \citep[e.g.][]{wang_etal_2021, vancap_etal_2025, lawlor_etal_2026}. Emission at these wavelengths comes from the planet's atmosphere, which for young gas giants is characterized by temperatures of $\sim$1000-2000 K. The atmospheres of protoplanets are commonly modeled as those of isolated objects cooling into empty space, utilizing standard substellar atmosphere grids intended for mature planets \citep[][]{husser_etal_2013, baraffe_etal_2015, marley_etal_2021}. But that neglects how ongoing accretion heats the atmosphere. Just as the atmospheres of hot Jupiters are heated externally by the intense radiation from their host stars \citep[e.g.][]{guillot_2010}, a protoplanet's atmosphere is heated from above by its accretion shock, which acts like a hot boundary layer. 

In this paper, we try to bridge the gap between the hot accretion shock and the underlying atmosphere. Our calculation is similar in spirit to that of \cite{berardo_etal_2017}, who also try to more smoothly connect these two regions. But where they focus on the interior entropy relevant to evolutionary models, we focus on calculating the structure of a protoplanet's atmosphere and the emission that it generates. We will show that accretion can restructure the atmosphere in ways that are measurable via emission spectroscopy in the infrared, which detects molecular absorption features that trace the various depths where they become transparent \citep[e.g.][]{molliere_etal_2019}. 

Section \ref{sec:model} describes how we model the impact of accretion on a planet's atmosphere. Section \ref{sec:results} presents results for a fiducial case, and then surveys how results change across parameter space. We summarize and discuss in Section \ref{sec:summary}.

\begin{table*}
\centering
\begin{tabular}{ccc}
\hline
\\[-2mm]
Parameter Name &  Description & Value(s) \\
\hline
$\Mp$ & Planet mass & 5$\Mj$ \\
$\Rp$ & Planet radius & 2$\Rj$ \\
$T_{\rm int}$ & Planet's effective temperature without accretion & 1500 K \\
$\Mdot$ & Planet's mass accretion rate & 
$ \leq 0.3\,\Mj/\mathrm{Myr}\,\,[\leq 3 \times 10^{-10}\,\Msun/\mathrm{yr}]$
\\
$f_{\rm down}$ & Fraction of downward-traveling shock radiation & \{0.1, 0.5, 1\} \\
$\kappa_{\rm shock}$ & Opacity to shock radiation & $\{1,\,10,\,100\}\,\rm cm^2/g $\\
$f_{\rm fill}$ & Covering fraction of accretion columns & \{0.1, 0.5,\, 1\} \\
\\[-2mm]
\hline
\hline
\end{tabular}
\caption{Summary of model parameters. Section \ref{subsec:fiducial} presents results for a fiducial model with $\{f_{\rm down},\,\kappa_{\rm shock},\,f_{\rm fill}\} = \{0.5,\,10\,\mathrm{cm^2/g},\,1\}$. Alternate parameters are explored in Section \ref{subsec:alternate_models}.}
\label{tab:summary}
\end{table*}
\section{Model for accretionally heated atmospheres}
\label{sec:model}
A protoplanet's atmosphere is heated from below by its own luminosity, and from above by accretion. Our goal is to calculate the structure and emergent spectrum for such an atmosphere, modeled as one-dimensional and plane-parallel. Section \ref{subsec:setup} describes our physical picture. Section \ref{subsec:numerics} details how we implement that picture in the radiative transfer code \texttt{picaso} \citep{batalha_etal_2019, mukherjee_etal_2023, mang_etal_2026}.

\subsection{Setup}
\label{subsec:setup}
Our fiducial protoplanet has mass $\Mp = 5M_{\rm J}$ and photospheric radius $\Rp = 2 R_{\rm J}$. The corresponding gravitational acceleration in the atmosphere is $g = G\Mp/\Rp^2 \simeq 3000$ cm/s$^2$. These values are similar to those of the PDS 70 and WISPIT 2 protoplanets \citep{wang_etal_2021, vancap_etal_2025}.

The planet's Kelvin-Helmholtz contraction generates a luminosity in its deep interior. We use a fiducial value $L_{\rm int} = 2 \times 10^{-4}\,\Lsun$ for this internal power. Absent accretion, the planet's effective temperature would be $T_{\rm int} = \left[L_{\rm int}/(4\pi \Rp^2 \sigma_{\rm SB})\right]^{1/4} \simeq 1500$ K, where $\sigma_{\rm SB}$ is the Stefan-Boltzmann constant. These values are consistent with evolutionary tracks of passively cooling planets at ages of a few Myr \citep[e.g.][]{baraffe_etal_2015}.

The planet accretes gas at a rate $\Mdot$. We assume that it has entered the ``runaway'' phase of growth in which flows can accelerate to supersonic speeds \citep[e.g.][]{pollack_etal_1996}. But otherwise, our model is largely independent of the geometry and dynamics of the accretion flow on Bondi/Hill sphere scales \citep[][]{choksi_etal_2023}. Regardless of whether runaway accretion happens through a circumplanetary disk or a more spherical circumplanetary envelope, the flow shocks against the planet's atmosphere at a depth \citep[e.g.][]{hartmann_etal_2016}
\begin{align}
P_{\rm shock} &= \frac{1}{2}\rho_{\rm preshock}v_{\rm ff}^2 \nonumber \\
&= 8 \times 10^{-6}\,\mathrm{bar}
\left(\frac{\Mdot}{0.1\,M_{\rm J}/\mathrm{Myr}}\right)
\left(\frac{\Mp}{5M_{\rm J}}\right)^{1/2}\left(\frac{\Rp}{2\,R_{\rm J}}\right)^{-5/2} \nonumber \\
&\qquad\times
f_{\rm fill}^{-1},
\end{align}
where $\rho_{\rm preshock} = \Mdot/(4\pi \Rp^2 v_{\rm ff} f_{\rm fill})$ is the density of infalling gas, $v_{\rm ff} = \sqrt{G\Mp/\Rp}$ is the free-fall velocity onto the planet's surface, and $G$ is the gravitational constant.\footnote{Technically, the distance from the planet center to the shock exceeds the radius $\Rp$ of the continuum photosphere. But this correction is unimportant because pressure falls exponentially with altitude.} 
We allow for non-spherical inflow by incorporating the parameter $f_{\rm fill} \leq 1$, equal to the fraction of the planet's surface covered by accreting columns. We assume the atmosphere below the shock obeys hydrostatic equilibrium.

A limitation of our study is that we only consider $\Mdot \lesssim 0.3\,\Mj/\rm Myr$. We make this restriction because our \texttt{picaso} simulations do not converge for higher accretion rates. The corresponding mass doubling times $\Mp/\Mdot$ exceed disk lifetimes of a few Myr, so this range is associated with the later stages of planet formation. Fortunately, our modeled range is still relevant to most known accretors. The accretion rates of the PDS 70 and WISPIT planets have been suggested to be $\Mdot \sim (0.01-1)\,\Mj/\rm Myr$ \citep[e.g.][]{haffert_etal_2019, choksi_chiang_2022}. Our modeled range is also applicable to the many brown dwarfs showing signs of residual accretion at rates $\Mdot \ll 1\,\Mj/\rm Myr$ \citep[e.g.][]{betti_etal_2023, zhang_etal_2024, cugno_etal_2024, finley_etal_2026}.

The shocked gas cools by emitting a variety of hydrogen-line and recombination-continuum photons \citep{aoyama_etal_2020, aoyama_etal_2021}. A fraction of that radiation escapes into space. Our focus is on the remaining fraction $f_{\rm down}$, which heats the atmosphere underneath. The irradiating flux is
\begin{align}
F_{\rm shock} &= \frac{1}{2}\rho_{\rm preshock}v_{\rm ff}^3 f_{\rm down} \nonumber \\
&= 3 \times 10^7\,\frac{\mathrm{erg}}{\rm sec\,cm^2}
\left(\frac{\Mdot}{0.1\,\Mj/\mathrm{Myr}}\right)
\left(\frac{\Mp}{5\,\Mj}\right) \nonumber \\
&\qquad\times
\left(\frac{\Rp}{2\,R_{\rm J}}\right)^{-3}
\left(\frac{f_{\rm down}}{0.5}\right) f_{\rm fill}^{-1}.
\end{align}
Our fiducial $f_{\rm down} = 0.5$ follows the naive expectation that about half the radiation from the shock heats the planet (\citealt{stahler_etal_1980}; \citealt{calvet_gullbring_1998}).
Radiation-hydrodynamic simulations of accretion shocks report $f_{\rm down} \approx 0.3-0.8$ (equation A4 of \citealt{aoyama_etal_2021}; see also \citealt{marleau_etal_2017} and \citealt{gottstein_etal_2026}). The limit $f_{\rm down} \ll 1$ guarantees a classic ``cold start'' \citep[e.g.][]{marley_etal_2007} in which nearly all of the accretional energy escapes into space via shock-cooling lines without heating the planet underneath. Conversely, ``hot start'' initial conditions require larger values of $f_{\rm down}$ so that much of the accretion power is directed into the planet. We will test $f_{\rm down} = \{0.1, 0.5, 1\}$.

The radiation emanating from the shock is absorbed by the layers underneath. The atmosphere's opacity to this radiation depends on the kind of photons produced in the shock. For example, if the shock emits mostly in Ly$\alpha$ (as suggested by \citealt{aoyama_etal_2020}), then photodissociation of water and methane could boost the opacity up to $\sim$1000$\,\rm cm^2/\rm g$ \citep[][]{heays_etal_2017}. If instead the shock were to radiate mostly H$\alpha$, the incoming photons would see a continuum opacity from pressure-broadened line wings that could be a few orders of magnitude smaller \citep{freedman_etal_2014}. Since the spectrum of radiation emerging from the shock is uncertain, we introduce a free parameter $\kappa_{\rm shock}$. We understand this parameter as the atmosphere's opacity averaged over the irradiating spectrum, whatever that may be. We test $\kappa_{\rm shock} = \{1,10,100\}\,\rm cm^2/g$. Table \ref{tab:summary} recaps our model parameters.

We highlight some limitations of our model. First, we neglect matter above the shock. If the preshock flow is optically thick, it could reprocess a fraction of the radiation back into the atmosphere. Accounting for this effect would only accentuate the accretional heating that we study. Second, our setup is 1D. Although we have partially accounted for non-spherical accretion geometries through the filling factor $f_{\rm fill}$, we do not account for the horizontal transport of heat that may occur between accreting columns and their cooler surroundings.

\subsection{Radiative transfer with \texttt{picaso}}
\label{subsec:numerics}
We use the 
\texttt{picaso} package to compute the thermal structure of the atmosphere sketched out above \citep{batalha_etal_2019}. Our main tool is the package's ``climate'' module \citep{mukherjee_etal_2023}, which iteratively solves for the pressure-temperature profile of a 1D column of atmosphere in radiative-convective-thermochemical equilibrium. To do so, the code integrates the two-stream, frequency-dependent equations of radiative transfer. Opacities are drawn from a self-consistent calculation of the equilibrium chemistry (more on this below). The solution is anchored by the planet's internal temperature. Absent accretion, the emergent flux from the atmosphere is 
$\sigma_{\rm SB}T_{\rm int}^4$.

The \texttt{picaso} solver operates on a grid of pressures. Our model atmosphere has $N_{\rm grid} = 110$ levels separated uniformly in log-space from $P_{\rm min} = 10^{-8}$ bar to $P_{\rm max} = 10$ bar. This range is large enough that our grid always encompasses the accretion shock towards its top and the radiative-convective boundary towards its bottom. We fix the number of convective zones to one. 

The shock irradiates the layers underneath it. To account for this heating, we deposit extra energy into grid cells with $P > P_{\rm shock}$ using the \texttt{EnergyInjection} module \citep{mang_etal_2026}. This module takes as input an extra energy flux $\Delta F_i$ to be deposited in every grid cell. The solver then iterates until the net upward flux from the $i$-th layer equals $\sigma_{\rm SB} T_{\rm int}^4$ plus the cumulative flux injected below that layer, ensuring the atmosphere is in steady-state. In our case, the extra energy is deposited by radiation, so the flux added to the $i-$th cell is 
\begin{align}
\Delta F_i &= F_{\rm shock}\left(e^{-\tau_{\rm shock, i-1}} - e^{-\tau_{\rm shock, i}}\right).
\end{align}
The optical depth to the shock radiation at each level follows from hydrostatic equilibrium:
\begin{align}
\tau_{\rm shock, i} &= \frac{\kappa_{\rm shock}}{g}\left(P_i - P_{\rm shock}\right).
\end{align}

The opacity $\kappa_{\rm shock}$, which depends on the unknown spectrum of shock radiation, is only used in depositing external heat. In computing the thermal response of the atmosphere, we use opacities based on equilibrium chemical abundances (see section 2.1.3 of \citealt{mukherjee_etal_2023} for details). These opacities are updated iteratively so that they are self-consistent with the pressure and temperature in each layer. 
 Numerically, \texttt{picaso} uses the ``correlated-k'' approach and divides opacities into 196 wavelength bins \citep[e.g.][]{marley_mckay_1999}. For our fiducial runs, we neglect the opacity from clouds. We consider cloudy atmospheres in Section \ref{subsubsec:clouds}.

The \texttt{picaso} solver requires an initial guess of the atmosphere's pressure-temperature profile. We used a warm-start approach to speed up the time to convergence. To begin, we ran a simulation of weak accretion $\Mdot = 10^{-3}\,\Mj/\rm Myr$. For this run, we used as a guess the profile of a non-accreting atmosphere drawn from the Sonora-Bobcat models \citep{marley_etal_2021}, extended isothermally above the top of that grid. Then we feed the converged solution so obtained as a guess for a new run with a larger $\Mdot$. Our models typically converge in $\sim$10 iterations. Appendix \ref{sec:convergence} confirms the robustness of our results against various numerical choices.

We postprocess our converged atmospheric structures to calculate the emergent spectrum. To do so, we apply an opacity table\footnote{\url{https://zenodo.org/records/3759675}} sampled at $R = 10,000$ to the pressure-temperature-abundance profile. We then bin down to a spectral resolution $R = 1000$, comparable to that achieved by \textit{JWST} NIRSpec.


\section{Results}
\label{sec:results}
We first present results for a fiducial model. The planet's properties are $\{\Mp,\,\Rp,\,T_{\rm int}\} = \{5\,\Mj,\,2\,\Rj,\,1500\,K\}$. Other parameters related to the accretion flow are the opacity to shock radiation $\kappa_{\rm shock} = 10\,\rm$cm$^2$/g, the fraction of shock radiation that heats the planet $f_{\rm down} = 0.5$, and the filling factor of accretion columns $f_{\rm fill} = 1$. In Section \ref{subsec:alternate_models} we test alternate models.

\subsection{Fiducial model}
\label{subsec:fiducial}

\begin{figure}
\includegraphics[width=\columnwidth]{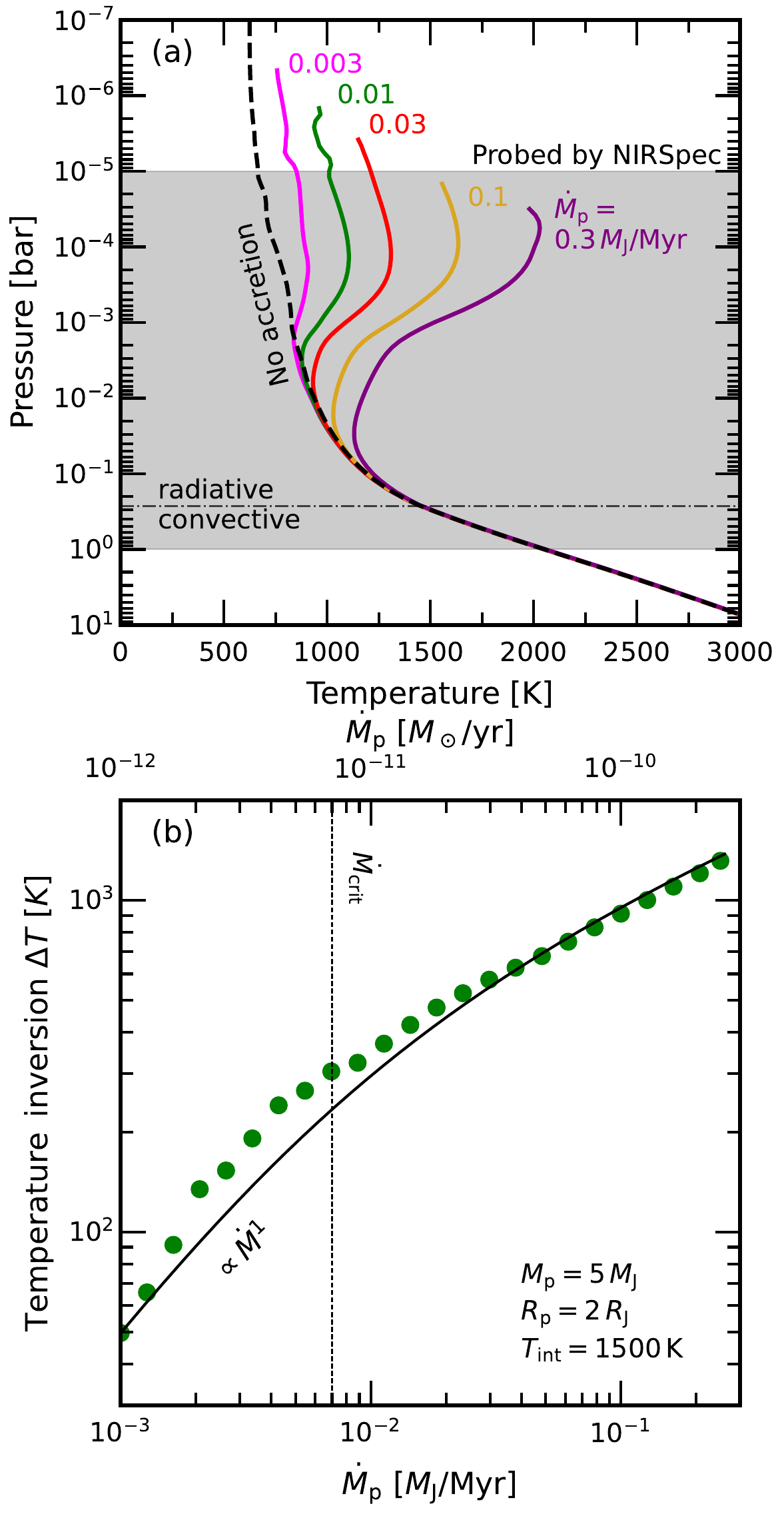}
\vspace{-6mm}
\caption{\textit{Top:} Pressure-temperature (P-T) profiles in the atmospheres of protoplanets. The black dashed curve shows the case without accretion.
Colored curves correspond to different accretion rates as labeled. Each curve terminates at the accretion shock. Irradiation from the shock drives a temperature inversion in the atmosphere. The gray shaded band highlights an approximate range of pressures traced by high/medium-resolution infrared spectroscopy (see also Figure \ref{fig:spec}). The horizontal line marks the radiative-convective boundary. 
\textit{Bottom:} The increase in atmospheric temperature $\Delta T$ vs.~accretion rate $\Mdot$. Points show the 
maximum temperature difference between an accreting and non-accreting atmosphere. The black curve plots the analytic prediction given by equation \ref{eqn:analytic_model}. The dashed vertical line marks $\dot{M}_{\rm crit}$, the center of the transition between the weak and strong-heating limits.
}
\label{fig:ptfid}
\end{figure}

Figure \ref{fig:ptfid}a plots the pressure-temperature (P-T) structure of our fiducial protoplanet accreting at different rates $\Mdot$. Absent accretion, the atmosphere has a familiar structure (black dashed curve). Below the radiative-convective boundary at $P_{\rm rcb} \approx 0.3$ bar it tracks d$\ln T / \mathrm{d}\ln P \approx 0.25$, the adiabat appropriate to molecular hydrogen with rotational degrees of freedom. Its upper layers, optically thin to continuum radiation, asymptote to a nearly isothermal ``skin'' of temperature $T_{\rm skin, 0} \approx 600$ K. 

Including accretion heats the atmosphere from above, creating temperature inversions (colored curves). Even the relatively low accretion rates that we model raise the atmosphere's temperature by hundreds of Kelvin. The heated layer is approximately isothermal and extends from $P_{\rm shock} \sim 10^{-6}-10^{-4}$ bar (depending on the accretion rate) down to $P \sim 10^{-3}$ bar. Such atmospheric pressures can be probed by high- or moderate-resolution infrared spectroscopy (gray shaded band; e.g.~\citealt{xuan_etal_2022, zhang_etal_2024, xuan_etal_2026}). Later in this section, we will explicitly calculate the emergent spectrum from these atmospheres. 

To summarize these simulation results, we record from each the temperature increase due to accretion. This $\Delta T$ is calculated by measuring the maximum temperature difference between a given P-T profile and that of the non-accreting atmosphere. Fig.~\ref{fig:ptfid}b plots how $\Delta T$ depends on $\Mdot$. We find that $\Delta T$ grows steeply with accretion rate for $\Mdot \lesssim 0.01\,\Mj/\rm Myr$, and more weakly above that threshold. 

To help understand these numerical results, we consider the Eddington two-stream solution for a double-gray atmosphere \citep[][]{hubeny_etal_2003, hansen_2008}. This model describes an atmosphere that is heated from below by internal radiation and from above by some external radiation field. The opacity to outgoing thermal radiation is a constant $\kappa_{\rm th}$, and it is smaller than the opacity to the shock radiation by a factor $\gamma = \kappa_{\rm shock}/\kappa_{\rm th} > 1$. By contrast, our \texttt{picaso} simulations use opacities that depend on wavelength and respond to ambient pressure and temperature.

In this analytic model, external heating establishes an isothermal layer at the top of the atmosphere whose temperature is given by (eq.~29 of \citealt{guillot_2010} with $\tau=0$):
\begin{equation}
T_{\rm skin}^4 = T_{\rm skin,0}^4 + \frac{\sqrt{3}}{4}\gamma\,T_{\rm acc}^4,
\label{eqn:skin}
\end{equation}
where 
\begin{align}
T_{\rm acc} &= \left(\frac{F_{\rm shock}}{\sigma_{\rm SB}}\right)^{1/4} \nonumber \\
&= 840\,\mathrm{K}
\left(\frac{\Mdot}{0.1\,\Mj/\mathrm{Myr}}\right)^{1/4}
\left(\frac{\Mp}{5\,\Mj}\right)^{1/4} \nonumber \\
&\qquad\times
\left(\frac{\Rp}{2\,R_{\rm J}}\right)^{-3/4}
\left(\frac{f_{\rm down}}{0.5}\right)^{1/4} f_{\rm fill}^{-1/4}
\label{eqn:Tacc}
\end{align}
measures the degree of accretional heating.\footnote{An expression similar to equation \ref{eqn:skin} is commonly used to model the atmospheres of  hot Jupiters \citep[e.g.][]{guillot_2010}. For that application, $T_{\rm acc}$ is replaced by a temperature $T_{\rm eq}$ that depends on the flux from the star at the planet's orbit.} To compare to our simulation data in Fig.~\ref{fig:ptfid}b, we calculate the temperature difference $\Delta T = T_{\rm skin} - T_{\rm skin,0}$ due to accretion. Strictly speaking, this only gives the temperature difference at the top of the atmosphere. But since the heated layer is nearly isothermal, exactly where we measure $\Delta T$ matters little. Inserting this definition into equation \ref{eqn:skin}, we obtain 
\begin{align}
\Delta T &= T_{\rm skin,0}\left[\left(1 + \frac{\Mdot}{\dot{M}_{\rm crit}}\right)^{1/4} - 1\right].
\label{eqn:analytic_model}
\end{align}
The critical accretion rate is 
\begin{align}
\dot{M}_{\rm crit} 
&\simeq 0.007\,\Mj\,\mathrm{Myr}^{-1}
\left(\frac{\gamma}{10}\right)^{-1}
\left(\frac{T_{\rm skin,0}}{600\,\rm K}\right)^{4},
\end{align}
defined as the rate for which the external heating term $\gamma T_{\rm acc}^4$ in equation \ref{eqn:skin} equals the internal term $T_{\rm skin,0}^4$. In this expression, we used values motivated by our \texttt{picaso} runs. First, the opacity ratio $\gamma = 10$ is calculated using the Planck mean of opacities at the top of the modeled atmosphere. Second, we used $T_{\rm skin,0} = 600$ K as found in our run without accretion.\footnote{An alternative choice would have been to use $T_{\rm skin,0} = T_{\rm int}/2^{1/4} = 1250$ K, the value appropriate for a strictly double-gray atmosphere. But our modeled atmospheres have colder skins than the double-gray solution predicts because they can cool via molecular bands \citep[e.g.][]{parmentier_guillot_2014}.  }
So normalized, equation \ref{eqn:analytic_model} does a good job at reproducing the numerical results shown in Fig.~\ref{fig:ptfid}b. The gradual shallowing in slope can be understood in terms of two limits. When $\Mdot \ll \dot{M}_{\rm crit}$, expanding equation \ref{eqn:analytic_model} to leading order gives $\Delta T \propto \Mdot^1$. In the more strongly heated limit, $\Delta T \propto \Mdot^{1/4}$. The transition between the two regimes happens slowly, on account of the 1/4 power in equation \ref{eqn:analytic_model}. Our simulations, which only extend to $\Mdot \sim 30\dot{M}_{\rm crit}$, do not fully realize the strongly-heated limit.

\begin{figure*}
\includegraphics[width=\textwidth]{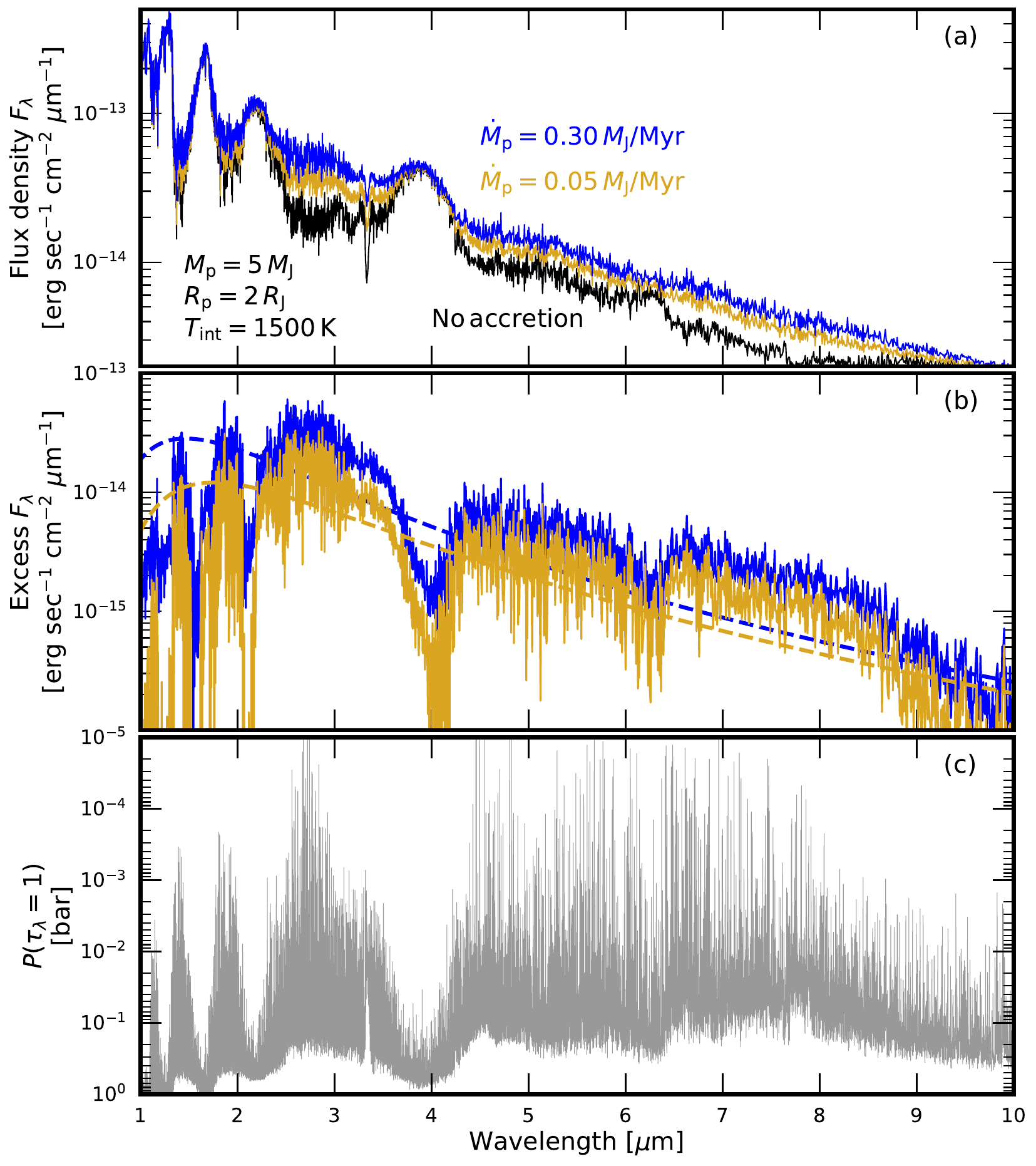}
\caption{Examples of how accretion modifies a planet's spectrum. The top panel plots spectra in flux density $F_{\lambda}$ vs.~wavelength $\lambda$ at a resolution $R = 1000$. Fluxes are normalized assuming the planet is 150 pc away from the observer, the typical distance of star-forming regions.
The spectrum of the atmosphere without accretion is shown in black. Colored curves show the spectrum corresponding to two different accretion rates as labeled. Accretional heating tends to mute absorption features by boosting the flux in molecular absorption bands.
The middle panel shows the excess flux due to accretion, obtained by differencing either of the colored curves and the black curve from the top panel. The dashed curves show the excess emission expected if the heated atmosphere radiated as an optically thin blackbody (equation \ref{eqn:dilute}). The bottom panel shows the pressure at which the atmosphere becomes optically thin to photons of a given wavelength $\lambda$ (optical depth $\tau_{\lambda} = 1$). By comparing to the upper two panels, we see that the excess emission concentrates in wavebands whose emission surfaces trace the upper layers of the atmosphere where accretional energy is deposited.}
\label{fig:spec}
\end{figure*}

\begin{figure}
\includegraphics[width=\columnwidth]{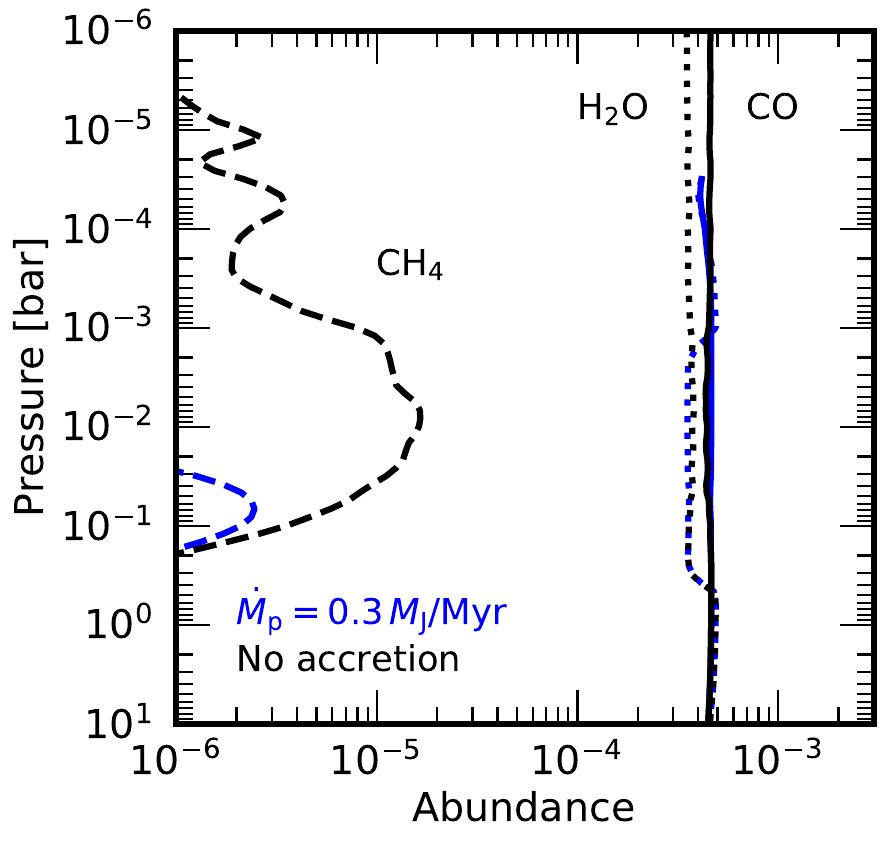}
\caption{How accretion can change the chemical makeup of a planet's atmosphere. The black curves show equilibrium molecular abundance profiles in our fiducial, non-accreting atmosphere. Different linestyles correspond to different molecules as labeled. Blue curves show results for $\Mdot = 0.3\,\Mj/\rm Myr$. Accretional heating destroys methane in the upper atmosphere (CH$_4$; dashed curve) and processes carbon into CO instead (solid curve). }
\label{fig:chemistry}
\end{figure}

Figure \ref{fig:spec}a plots emergent spectra for different accretion rates. Accretional heating makes the atmosphere brighter. If the atmosphere were strictly double-gray, we would expect the excess emission to match that of a dilute blackbody:
\begin{align}
F_{\lambda,\,\rm excess} \sim \frac{F_{\rm shock}}{\sigma_{\rm SB} T_{\rm skin}^4} \times B_{\lambda}(T_{\rm skin}),
\label{eqn:dilute}
\end{align}
where $B_{\lambda}$ is the Planck function and the ratio $F_{\rm shock}/\sigma_{\rm SB} T_{\rm skin}^4$ is an estimate for the heated layer's optical depth to its own thermal emission. Figure \ref{fig:spec}b shows that equation \ref{eqn:dilute} only roughly traces the measured excess. In our simulations, the excess emission concentrates in molecular bands that radiate most strongly in the upper atmosphere where accretion power is deposited. The most prominent excesses are found in the 2.5-3.5 $\mu$m and 4.5-9 $\mu$m ranges. The former band is covered by the $\nu_1$ and $\nu_3$ symmetric and asymmetric stretching modes of water. The latter includes the CO fundamental mode and the $\nu_2$ bending mode of water. Comparatively little excess emission is observed in opacity windows like 3.5-4.5 $\mu$m. Because the atmosphere is optically thinner at these wavelengths, the observed emission comes from deeper layers undisturbed by accretion (see Fig.~\ref{fig:spec}c). Thus accretional heating reduces the contrast between absorption bands and the surrounding continuum. For our modeled range $\Mdot \lesssim 0.3 \,M_{\rm J}/\rm Myr$, the flux in the bands can be boosted by a factor of $\sim$3. Even higher accretion rates might reverse absorption features into emission.



By heating the atmosphere, accretion also alters its chemical makeup. Figure \ref{fig:chemistry} plots abundances profiles of a few different molecules in our simulations. We see that accretion destroys practically all the CH$_4$ in the heated layer, as the higher temperatures process carbon into CO. This change can also be made out in the spectra plotted in Fig.~\ref{fig:spec}, which shows a shallowing of the 3.3 $\mu$m band, corresponding to the asymmetric bending mode of CH$_4$, as accretion rate increases.

\begin{figure}
\includegraphics[width=\columnwidth]{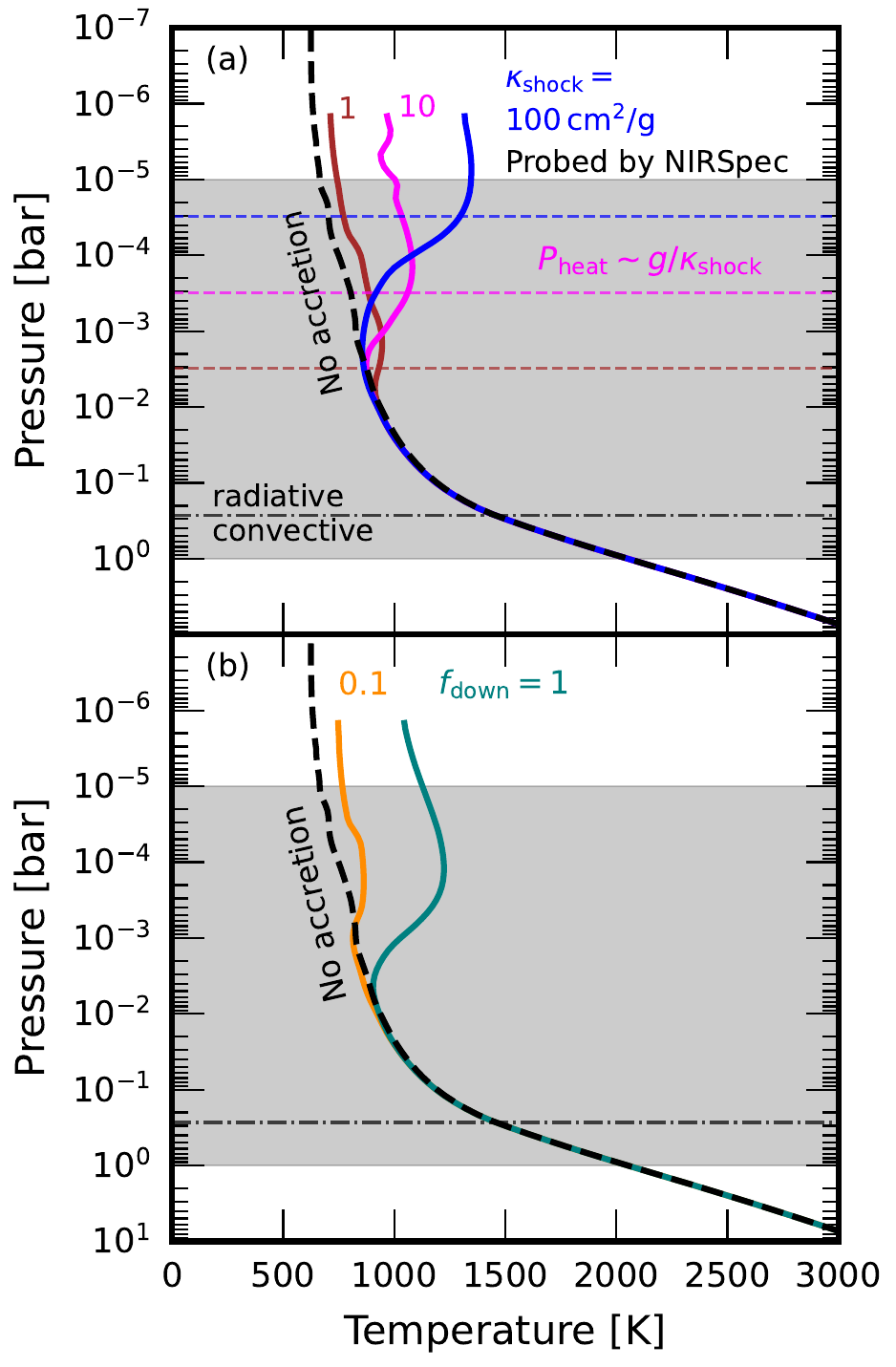}
\caption{Similar to Fig.~\ref{fig:ptfid}a, but now showing how the atmosphere responds to changes in model parameters at fixed $\Mdot = 0.01\,\Mj/\rm Myr$. The top panel varies the atmosphere's opacity to shock radiation $\kappa_{\rm shock}$. Larger values concentrate the accretional energy into a narrower layer. Dashed lines mark an order-of-magnitude estimate for the depth of the heated layer. The bottom panel varies $f_{\rm down}$, the fraction of accretion power that contributes to atmospheric heating.
}
\label{fig:params}
\end{figure}

\begin{figure*}
\includegraphics[width=\textwidth]{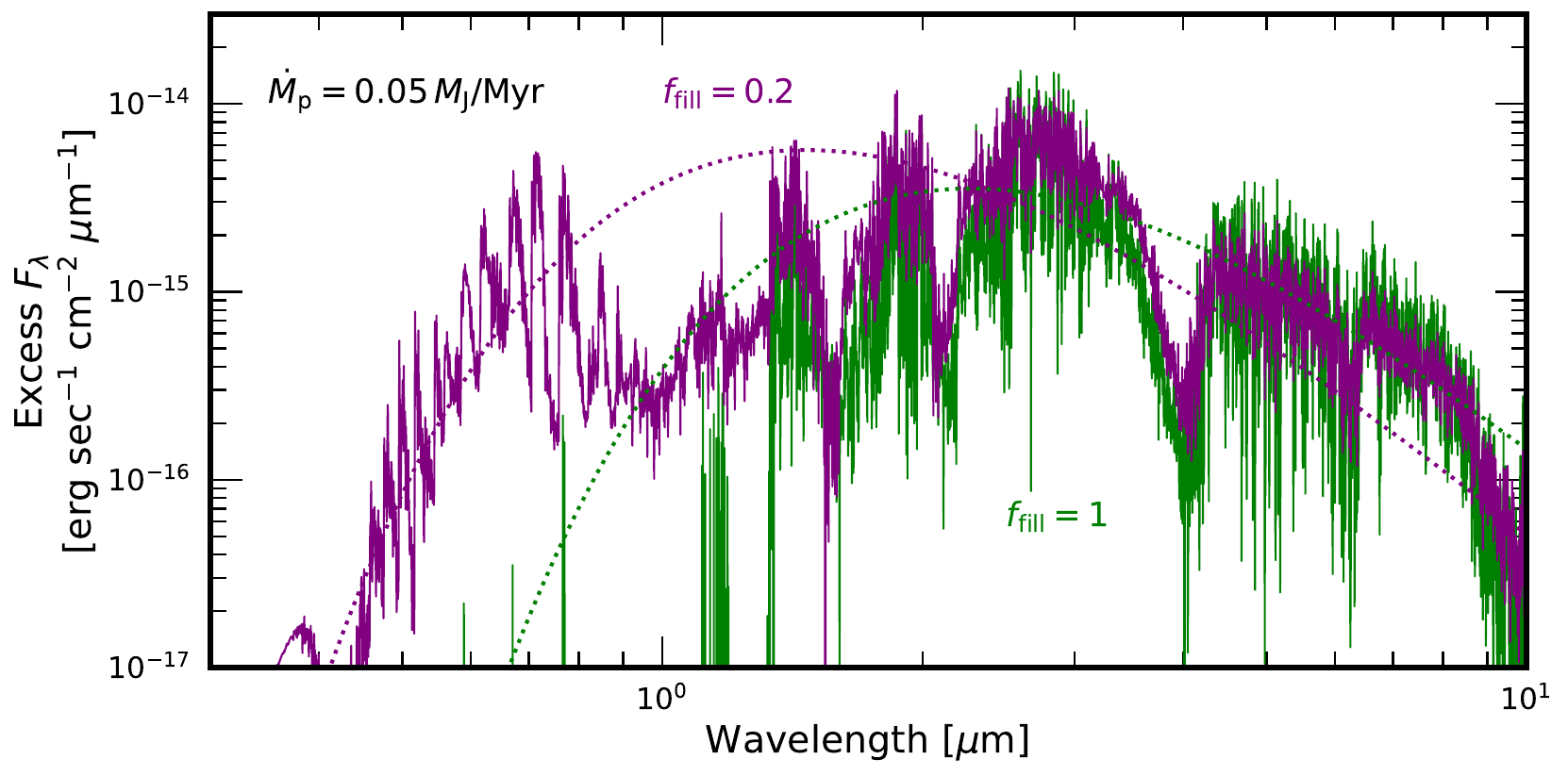}
\caption{Effect of accretion geometry on a protoplanet's spectrum. The $y$-axis plots the excess flux relative to the undisturbed atmosphere. The green curve shows the case where the planet accretes in a spherically symmetric manner at a rate $\Mdot = 0.05\,\rm \Mj/Myr$. The purple curve shows a case with the same $\Mdot$, but now assuming accretion only happens over a fraction $f_{\rm fill} = 0.2$ of the planet's surface. The resulting excess spectrum arises from a combination of accretional hotspots dominating the emission at short wavelengths plus a redder non-accreting component. Dotted lines indicate approximate excesses that would be expected if the accreting regions radiated as dilute blackbodies.
}
\label{fig:hotspot}
\end{figure*}

\begin{figure*}
\includegraphics[width=\textwidth]{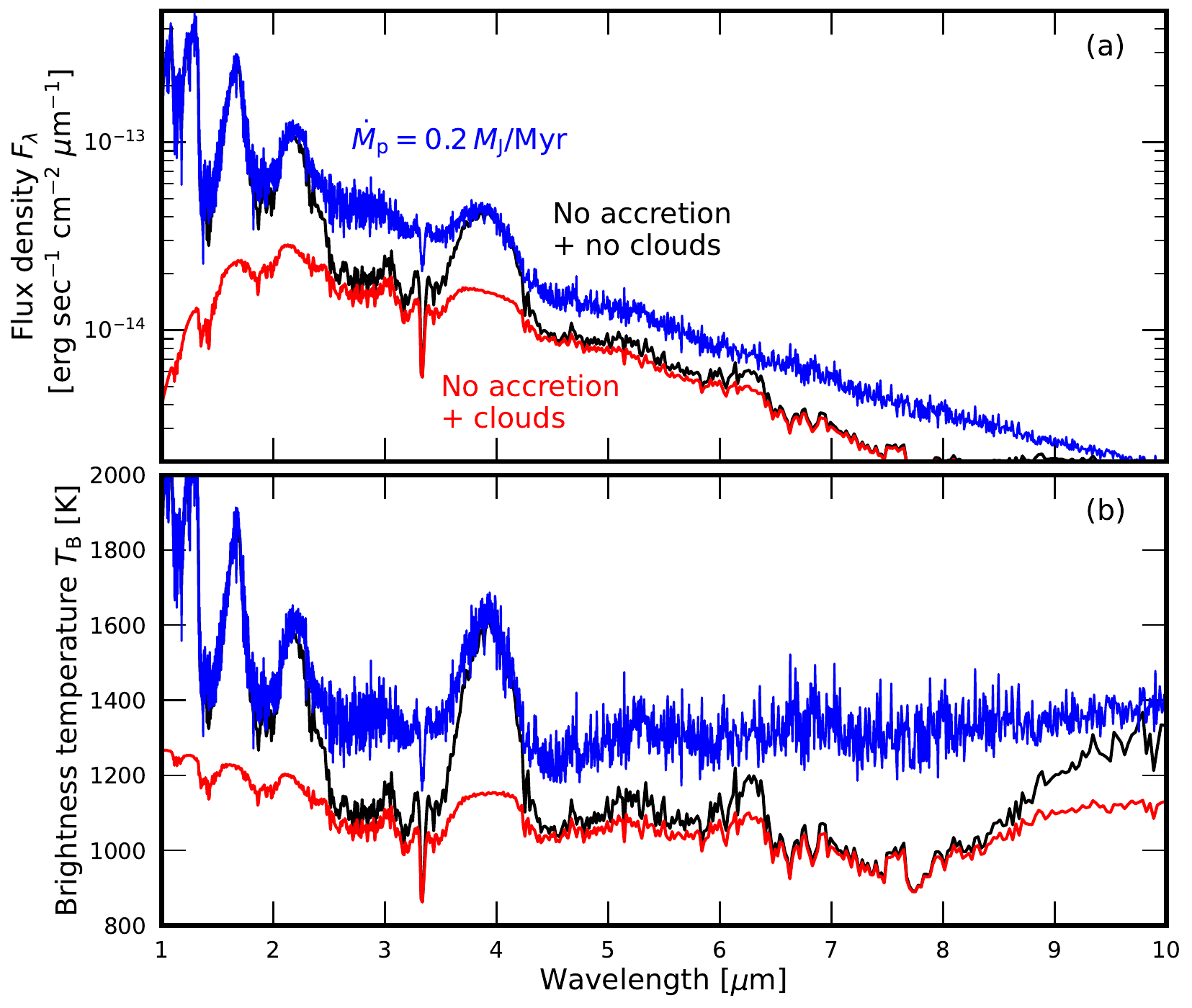}
\caption{How clouds affect the detectability of accretional heating. The blue curve in the top panel shows the spectrum of an accreting planet (same as in fig.~\ref{fig:spec}). The red curve shows a cloudy, non-accreting atmosphere. Cloud opacities are calculated from \texttt{virga} for silicate and iron condensates, assuming an eddy diffusivity $K_{\rm zz} = 10^{9}$ cm$^2$/s and sedimentation efficiency $f_{\rm sed} = 2$.
Compared to the non-accreting, non-cloudy baseline shown in black, both cloudy atmospheres and accretionally heated atmospheres exhibit weaker contrasts between absorption bands and the continuum in the near-infrared. The bottom panel further illustrates this point by plotting brightness temperature $T_{\rm B}$ vs.~wavelength for the three cases. Clouds and accretion both damp fluctuations in $T_{\rm B}$, implying a less featured spectrum. Clouds push the continuum photosphere up to higher altitudes where temperatures are lower; accretion heats the high-altitude layers where lines form.}
\label{fig:cloudy_spec}
\end{figure*}

\subsection{Model variations}
\label{subsec:alternate_models}
We explore variations on the fiducial model presented above. In Section \ref{subsubsec:params} we survey how results depend on the parameters characterizing the accretion flow  $\kappa_{\rm shock}$, $f_{\rm down}$, and $f_{\rm fill}$. We change one of these at a time while others are held fixed at their fiducial values. In Section \ref{subsubsec:clouds} we consider the effect of clouds.

\subsubsection{Parameter survey}
\label{subsubsec:params}
Figure \ref{fig:params}a shows how changing $\kappa_{\rm shock}$ affects the atmosphere's P-T structure. Larger values of $\kappa_{\rm shock}$ concentrate the accretional energy into a narrower layer. An estimate for the depth down to which the heating extends is
\begin{align}
P_{\rm heat} &\sim \frac{g}{\kappa_{\rm shock}} \nonumber \\ 
&\sim 3 \times 10^{-4}\,\mathrm{bar}\left(\frac{\Mp}{5\,\Mj}\right)\left(\frac{\Rp}{2\,\Rj}\right)^{-2}\left(\frac{\kappa_{\rm shock}}{10\,\rm cm^2/g}\right)^{-1}.
\label{eqn:pheat}
\end{align}
Fig.~\ref{fig:params}a shows that our simulations roughly track this estimate (horizontal dashed lines). We note that external heating can stall a planet's cooling if the irradiating flux penetrates down to the planet's radiative-convective boundary \citep[e.g.][]{arras_bildsten_2006, ginzburg_quataert_2021}. For our modeled planet, that boundary sits at $P_{\rm rcb} \approx 0.3$ bar, well below the heated layer (black horizontal line). Thus the emission from the heated layer and the passively cooling planet below can largely be separated.

Fig.~\ref{fig:params}b adjusts the fraction of accretion power that is directed downward $f_{\rm down}$. In our model, this parameter and $\Mdot$ have similar effects on the final P-T structure, since the irradiating flux $F_{\rm shock}$ scales with their product. At a fixed accretion rate, reducing $f_{\rm down}$ deposits less heat into the atmosphere. In that case, correspondingly more power is assumed to escape into space in the form of nonthermal shock radiation.

Finally, we consider non-spherical accretion with $f_{\rm fill} < 1$. Concentrating the inflow into a fraction of the surface produces hot
accreting columns viewed against a backdrop of undisturbed atmosphere.
We construct an example of the resulting two-component spectrum by
weighting each component by the area it covers:
\begin{align}
F_{\lambda} = (1 - f_{\rm fill})F_{\lambda,\,\rm undisturbed}
+ f_{\rm fill}F_{\lambda}(\Mdot/f_{\rm fill}),
\end{align}
where $F_{\lambda,\,\rm undisturbed}$ is the spectrum of the
non-accreting atmosphere and $F_{\lambda}(\Mdot/f_{\rm fill})$ is the
spectrum of a spherically accreting model ($f_{\rm fill} = 1$) evaluated
at the rate $\Mdot/f_{\rm fill}$, so that the total accretion
rate onto the planet is $\Mdot$ in both cases. Figure \ref{fig:hotspot} shows that at fixed total accretion rate, a
planet accreting through hotspots appears bluer. The heated columns cover
less area but are hotter, shifting the excess emission to shorter
wavelengths.

\subsubsection{Clouds}
\label{subsubsec:clouds}
In this section we consider the effect of clouds. Our goal is not to model clouds in detail, but to get a sense for how they affect the detectability of accretional heating. To do so, we take a hybrid approach. We use the cloudless pressure-temperature structures that we have already computed and
postprocess them into spectra including opacities for silicate and iron condensates calculated from the \texttt{virga} cloud model \citep{ackerman_marley_2001, batalha_etal_2026, moran_etal_2025}. This approach neglects cloud feedback on the thermal structure but allows us to generate approximate spectra for a fraction of the computational cost (see \citealt{mang_etal_2026} for self-consistent calculations of P-T profiles in cloudy atmospheres).

Figure \ref{fig:cloudy_spec}a shows that the spectrum of a cloudy  atmosphere has some similarities to that of an accreting atmosphere. Both clouds and accretion mute absorption features. For example, in our ``control'' case of a non-cloudy, non-accreting atmosphere, the flux in the 2.5-3$\mu$m absorbing band is $\sim$2$\times$ lower than just outside the band at 4$\mu$m, compared to a flux contrast of $\lesssim 30\%$ for the cloudy or accreting atmospheres. Clouds effect this change by depressing the near-infrared continuum, whereas accretion boosts the flux in the absorbing bands. Figure \ref{fig:cloudy_spec}b helps interpret these changes by plotting the brightness temperature $T_{\rm B}(\lambda)$, as a proxy for the temperature of the material dominating emission at a given wavelength. Clouds push the continuum photosphere up to higher altitude, lower temperature environments; accretion heats the high-altitude layers where lines form. Both changes reduce the temperature contrast between the continuum and line-forming regions. Thus an accreting atmosphere might be confused with a cloudy atmosphere. 

Because we only add clouds in postprocessing, without accounting for their effect on the P-T profile, our cloudy spectra integrate to less than $\sigma_{\rm SB} T_{\rm int}^4$. In a self-consistent treatment, cloud back-warming would result in excess emission at wavelengths longer than a few micron that would compensate for the depression in flux at shorter wavelengths \citep[e.g.][]{marley_robinson_2015}. 
Our conclusion that absorption features would be muted should be robust in the optical/near-IR where cloud feedback is minimal.

\section{Summary and Outlook}
\label{sec:summary}
Observations of accreting planets are providing a real-time look at the planet formation process. To help interpret the spectra of these protoplanets, we have studied the structure of their atmospheres. The atmosphere is heated from below by the planet's cooling radiation and from above by nonthermal radiation generated in an accretion shock. This sets up a thermal inversion. Even the relatively weak accretion rates we studied, corresponding to mass doubling times exceeding system ages, can raise temperatures by many hundreds of Kelvin at ambient pressures of order millibars.  The heated atmosphere produces an optical/infrared excess. For the dust-poor environments we studied, the excess is carried by molecular lines that trace the upper atmosphere. An accretionally heated atmosphere may thus be distinguished by its muted absorption features. 
 
Interestingly, this description recalls the appearance of the protoplanets PDS 70 b\&c, both of which have been noted for their particularly shallow K-band absorption features compared to non-accreting counterparts with similar temperature and mass \citep{wang_etal_2020, wang_etal_2021}. Another interpretation that we cannot rule out is that the weak absorption features are due to a thick layer of clouds. Detection of condensate spectral features like the 10 $\mu$m silicate resonance might help pin down the relative importance of accretion and clouds. 

Because protoplanets do not conform to the pressure-temperature profiles of 
passively cooling objects, standard substellar spectral templates such as PHOENIX \citep{husser_etal_2013}, BT-Settl \citep{baraffe_etal_2015}, or Sonora-Bobcat \citep{marley_etal_2021} should not be used to fit their emission. Instead, we advocate for the use of ``free’’ retrievals of atmospheric structure that can accommodate non-standard pressure-temperature profiles \citep[e.g.][]{molliere_etal_2019}. In the system YSES 1, this approach has delivered perhaps the clearest evidence so far of accretional heating. Orbiting a Sun-like star, the $\sim$30 Myr-old brown dwarf YSES 1b exhibits H$\alpha$ emission indicative of ongoing accretion \citep{hoch_etal_2025}. \cite{zhang_etal_2024} leverage high-resolution K-band spectroscopy from VLT/CRIRES+ to reconstruct the structure of the emitting layers. They report YSES 1b’s atmosphere is hundreds of Kelvin hotter than expected for a passively cooling object (their fig.~5). 

More atmospheric retrievals on substellar accretors would be welcome. \textit{JWST} NIRSpec will soon deliver a medium-resolution ($R = 2700$) spectrum of the newly discovered WISPIT 2b that will enable the first retrieval on a bona fide protoplanet (PID 9993; PI: Balmer). The planetary-mass companion SR 12c and the brown dwarf GQ Lup b are also promising targets
\citep{finley_etal_2026, cugno_etal_2024}. For these objects, we suggest observing in shorter wavelength bands ($\lesssim 4 \mu$m) to minimize contamination from the circumplanetary disk (see \citealt{cugno_grant_2025} for an observed mid-IR spectrum of a CPD). 

In extreme cases, a thermal inversion can flip absorption features into emission. A molecular emission line has been observed in the Gyrs-old isolated brown dwarf W1935 \citep{faherty_etal_2024} and is thought to arise from a temperature inversion established by auroral heating  \citep{zuckerman_etal_2026, smith_etal_2026} or the breaking of gravity waves \citep{young_etal_1997}. Recently, \textit{JWST} has discovered methane emission from the $\sim$6$M_{\rm J}$ isolated object LRL 11001 \citep{luhman_etal_2026}. Given its membership in a star-forming complex and its age of just 2-3 Myr, LRL 11001 might be accreting. Detection of a circumplanetary disk or shock-tracing lines like H$\alpha$ would provide strong evidence that accretion powers the inversion.


Characterizing the atmospheres of protoplanets may help to distinguish between ``hot'' and ``cold'' start initial conditions of planet formation \citep{marley_etal_2007}. These two limits differ in the fraction of accretional energy that makes it past the shock \citep{marleau_etal_2017, berardo_etal_2017, aoyama_etal_2020}. 
We have tried to be agnostic about this question by varying the fraction $f_{\rm down}$ in our model. If planets form in a cold start, we expect their atmospheric structures to hew closely to those of passively cooling objects; if they form in a hot start, retrievals should uncover thermal inversions and significantly elevated atmospheric temperatures.

A test of our picture would be to look for simultaneous temporal variability across the spectrum. If such behavior were observed, it would favor our thinking that much of the accretion power is reprocessed out to longer wavelengths. \cite{almendros-abad_etal_2025} study accretion onto the free-floating $\sim 5-10 M_{\rm J}$ object Cha 1107-7626. Encouragingly, they report a months-long burst in the H$\alpha$ flux is accompanied by a burst of similar magnitude in the optical continuum. Another promising target for multi-band monitoring would be the PDS 70 system, where variability in the planetary H$\alpha$ fluxes has recently been reported \citep{zhou_etal_2025, close_etal_2025_pds}. Pinning down the timescale(s) of variability would also be informative. Long-timescale variability might be set by bursts in the accretion flow, while shorter-timescale variability might be caused by accretion hotspots rotating into and out of view.

There are also avenues for future theoretical study. In this paper, we focused on accretion luminosities $G\Mp\Mdot/\Rp$ that are less than or comparable to planet's internal luminosity $L_{\rm int}$. Future work could study the limit where external heating overpowers the planet's cooling luminosity. Hot Jupiters live in this regime, and their inflated sizes have been attributed to the excess heat deposited deep in their interiors \citep[e.g.][]{arras_bildsten_2006}. Another avenue is to consider dust brought in with the accretion flow. If the heated layer is dusty, we expect that it will reprocess the accretion power more uniformly across the spectrum rather than concentrating its emission in molecular features. We plan to use the Monte Carlo radiative transfer code \texttt{radmc3d} to study this case \citep{dullemond_etal_2012}. Finally, our calculations assumed equilibrium molecular abundances set by the ambient pressure and temperature. But the nonthermal radiation from the shock might drive photochemistry that would alter these abundances. Future work could apply photochemistry codes like \texttt{VULCAN} \citep{tsai2021vulcan} to study the atmospheres of accreting planets.

The physical processes we have studied are analogous to those thought to take place in protostars (see \citealt{hartmann_etal_2016} for a review). In that context, shocked gas reaches $T_{\rm shock} \sim m_{\rm p}v_{\rm esc}^2/k_{\rm B} \sim 10^6$ K and cools off X-rays ($k_{\rm B}$ is Boltzmann's constant and $m_{\rm p}$ is the proton mass).\footnote{This $T_{\rm shock}$ describes the temperature of the gas immediately after the shock, before it has had a chance to radiatively cool (see section A.2.2 of \citealt{aoyama_etal_2020} for more discussion on this point).} Most of this shock radiation is thought to be reprocessed by the atmosphere below to produce optical/UV continuum excesses that ``veil'' absorption features \citep[e.g.][]{calvet_gullbring_1998}. Our picture is an energetically scaled-down version of this scenario. The shallower potentials of protoplanets yield shock temperatures $\sim 10^5$ K, so H$\alpha$ and Ly$\alpha$ photons replace X-rays \citep{aoyama_etal_2020}.\footnote{T-Tauri stars also emit H$\alpha$. But the mechanism for the emission is thought to differ in the protostellar and protoplanetary contexts. In the former case, H$\alpha$ is thought to come from the preshock flow heated by the stellar magnetosphere \citep[e.g.][]{muzerolle_etal_1998}. In the latter, H$\alpha$ is thought to be generated in the shock itself \citep{aoyama_etal_2020, aoyama_etal_2026}.} As this shock-cooling radiation is reprocessed by the layers underneath, optical/infrared excesses take the place of UV/optical excesses.


\section*{Acknowledgements}
We thank Konstantin Batygin, Brendan Bowler, Eugene Chiang, Sivan Ginzburg, Heather Knutson, Yoram Lithwick, Gabriel-Dominique Marleau, Eliot Quataert, and Jason Wang for helpful and motivating discussions. We are grateful to the Heising-Simons foundation for hosting the summit that launched this project, and for their support of NC, PG, JX, and WB through the 51 Pegasi b fellowship.

\bibliographystyle{aasjournal}
\bibliography{planets_nick}
\appendix
\twocolumngrid
\section{Convergence tests}
\label{sec:convergence}
\begin{figure}
\includegraphics[width=\columnwidth]{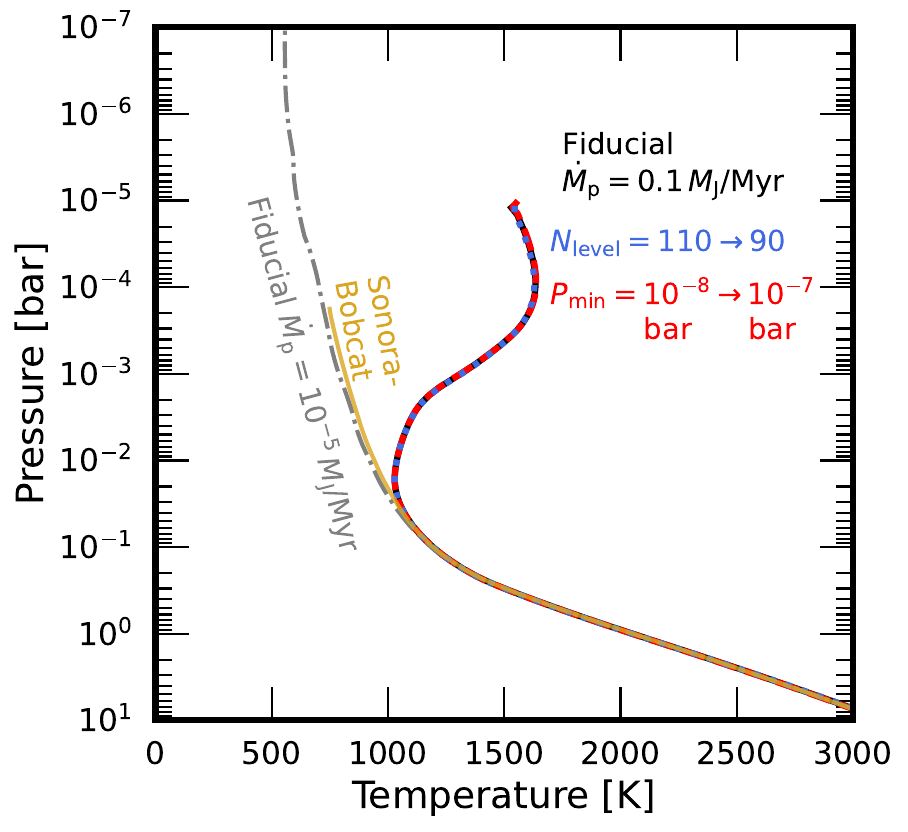}
\caption{Tests of numerical convergence. The solid black curve shows our fiducial model with $\Mdot = 0.1\,M_{\rm J}/\rm Myr$, 110 pressure levels, and a grid upper boundary of $P_{\rm min} = 10^{-8}$ bar. The dashed blue curve plots results for a run with fewer pressure layers $N_{\rm level} = 90$. The red dotted curve corresponds to a run whose upper boundary is moved to $P_{\rm min} = 10^{-7}$ bar. Neither of these choices affects the P-T profile noticeably. To check that our model relaxes to the correct structure in the limit of vanishingly small accretion rates, we plot the gray dash-dot curve which shows results for a run with our fiducial numerical choices and $\Mdot = 10^{-5}\,M_{\rm J}/\rm Myr$. This P-T structure compares well with the structure reported in the Sonora-Bobcat models of \citep{marley_etal_2021} at the pressure where results are available.}
\label{fig:convergence}
\end{figure}

Our \texttt{picaso} simulations were performed on a grid of pressures with $N_{\rm level} = 110$ layers uniformly separated in logspace between 10 bars and $P_{\rm min} = 10^{-8}$ bars. We perform three checks of our numerics. 
First, we try reducing the number of grid layers to $N_{\rm level} = 90$. Second, we try increasing the $P_{\rm min}$ of our models. Fig.~\ref{fig:convergence} shows that P-T profiles computed on these two grids are indistinguishable from those obtained in our fiducial model. As a third and final check, we verified that in the limit of very small but nonzero accretion rates, our models converge to previously reported results. Figure \ref{fig:convergence} compares our model results for $\Mdot = 10^{-5}\,M_{\rm J}/\rm Myr$ with those of the Sonora-Bobcat grids and finds good agreement.

\end{document}